\documentclass[onecolumn,useAMS]{mn2e}
\usepackage{epsfig}
\usepackage{graphicx}
\usepackage{dcolumn}
\usepackage{bm}
\def\la{\lower.5ex\hbox{$\; \buildrel < \over \sim \;$}}
\def\ga{\lower.5ex\hbox{$\; \buildrel > \over \sim \;$}}
\def\apj{ApJ}
\def\mnras{MNRAS}
\def\prd{PRD}
\def\apjl{ApJL}

\def\aj{AJ}

\def\physrep{Physics Reports}

\def\rss {\rm \scriptscriptstyle}

\begin{document}
\title[Generation of  Magnetic fields]{Generation of  Magnetic
Field in the Pre-recombination Era}
\author[Rajesh Gopal and Shiv K. Sethi]
{Rajesh Gopal$^1$ and Shiv K. Sethi$^1$   \\
\hspace{-0.1cm} ${}^1$Raman Research Institute, Bangalore 560080, India \\
\hspace{-0.1cm} emails: rajesh@rri.res.in, sethi@rri.res.in
}

\maketitle

\begin{abstract}
We study the possibility of generating magnetic fields during the evolution
of electron, proton, and photon plasma in the pre-recombination era. We show 
that a small magnetic field can be generated in the second order of 
perturbation theory for scalar modes with adiabatic initial conditions. 
The amplitude of the field is $\la 10^{-30} \, \rm G$ at the present epoch 
for scales from sub-kpc to $\ga 100 \, \rm Mpc$. 
\end{abstract}

\section{Introduction}
Magnetic fields are ubiquitous in the universe and  presumably 
play an important role  in most  objects in the universe. Their origin 
however is not well understood (see e.g. Parker 1979; Zeldovich, Ruzmaikin \& Sokoloff 1983). Galactic fields of micro-Gauss strength could have arisen 
from dynamo amplification of seed fields $\simeq 10^{-20} \, \rm G$ (see e.g. Ruzmaikin, Shukurov \& Sokoloff 1988; Beck et al 1996; Shukurov 2004;
Brandenburg \& Subramanian 2004). On the other hand, large scale magnetic fields could  have originated from primordial magnetic fields $\simeq 10^{-9} \, \rm G$
generated during inflationary epoch in the early universe 
(Turner \& Widrow 1988, Ratra 1992, see   Grasso \& Rubenstein 2001;
Giovannini 2004 for reviews).

 Small seed fields  could have arisen from various astrophysical processes 
during the later stages of evolution of the universe ($z\,\la\,20$) 
(e.g.  Subramanian, Narasimha and Chitre 1995; Kulsrud et 
al 1997; Grasso \& Rubenstein 2001 and Widrow 2002 for reviews). Harrison 
(1970) considered a scenario in which a small seed field 
$\simeq 10^{-25} \, \rm G$ is generated owing to the vorticity in the 
photon-baryon fluid in the pre-recombination plasma. More recently, 
Hogan (2000) and Berezhiani  \& Dolgov (2003) also considered
the generation of magnetic fields from photon pressure in the 
pre-recombination epoch. From astrophysical processes, during the post-
recombination universe, the typical scales of the seed field are 
$\simeq \hbox{a few} \rm Mpc$ (see e.g. Subramanian, 
Narasimha and Chitre 1995; Kulsrud et  al 1997).  On the other hand, magnetic 
fields can be generated on much larger scales in the pre-recombination 
universe. 

In the pre-recombination universe,  photons, electrons, and protons can 
be treated as tightly coupled fluids.   Photons however preferentially 
exert pressure on electrons (the pressure on protons is suppressed by 
a factor $(m_e/m_p)^2$). During the evolution of the 
photon-baryon plasma  a difference in velocity fields
of electrons and protons can therefore be generated, and 
 this  holds the promise of generating 
magnetic fields from this induced current.  In addition the approximation
that photons and baryons are tightly coupled, and therefore 
can be treated as one fluid,  breaks down for scales up to several $\rm Mpc$
by the time of recombination (during the process of recombination this 
scale  exceeds  the horizon scale). At smaller   scales
the photons free-stream and this in principle can lead to
 additional contribution
to induced currents that might generate  magnetic fields. 

In this paper we study 
the coupled   electron, proton, and photon plasma
   in first and second order in perturbation theory to understand the 
generation of magnetic fields during the evolution of the plasma. 

In the next section we describe the relevant equations for our study. 
In \S 3 we discuss the evolution of magnetic fields and its sources in the 
first and second order in perturbation theory. In \S 4 discuss our results 
and give concluding remarks. Throughout the paper, numerical values 
of different quantities  are given
for the spatially  flat FRW  model with  $\Omega_m = 0.3$ and $\Omega_\Lambda = 0.7$ (Spergel et al. 2003, Reiss et al. 2004, Tonry et al. 2003, Perlmutter  et al. 1999,
Riess  et al. 1998) with  $\Omega_b h^2 = 0.02$ (Spergel et al. 2003, Tytler 
 et al. 2000) and 
$h = 0.7$ (Freedman  et al. 2001).

\section{Pre-recombination plasma}
The primary components of the plasma in the radiation era which lasts between 
neutrino decoupling and recombination are 
photons, free electrons and protons. Observations of the Cosmic Microwave Background Radiation (CMBR) which is 
the relic of the radiation existing in this era show that the plasma 
is almost homogeneous and in thermal equilibrium at the time of recombination
(see e.g. Peebles 1993). However, anisotropies observed in the CMBR also
 indicate that there are spatial fluctuations superimposed on this uniform
 background density. The initial condition for each mode of fluctuation
 is set before it enters the horizon. In the present analysis we assume 
that the fluctuations are adiabatic which means that the entropy per fluid 
particle is conserved. Recent WMAP  observations favour this initial condition (Peiris et~al. 2003). The electrons interact with each other and with protons
 through Coulomb scattering. The mean free paths for $e\hbox{-}e$, $e\hbox{-}p$ and $p\hbox{-}p$  collisions are the same in this thermal plasma (see e.g. Shu 1992) and 
 are much smaller than the astrophysically relevant scales
 ($\simeq 1\, \rm Mpc$). Hence a continuum description treating them as 
 fluids can be used. In such a macroscopic description
 the effect of scattering between different species is taken into account 
by including a momentum exchange term in the Euler equation. For photons 
however the dominant interaction is Thomson scattering off free electrons 
with mean free path (comoving), $l_{\gamma e}$, at $z \simeq 1000$ 
for a fully ionized universe being, $l_{\gamma e}=1/(a\sigma_{\rss T}
n_{e}) \simeq  3 \, \rm Mpc$. 
Here, $\sigma_{\rss T}$ is the Thomson cross section for $e\hbox{-}\gamma$ 
scattering, $n_{e}$ is the electron number density and $a$ is the scale 
factor. This is comparable to the length scales in consideration and hence 
a Boltzmann particle description is essential.

\subsection{Description of photons}
The photons are described by the phase space distribution function 
\( f({\mathbf{x}},\eta,p,\hat{n}) \) where $p$ is the magnitude of photon 
momentum, \(\hat{n}\) is the propagation direction and $\eta$ is the 
conformal time. Since the distribution is blackbody to zeroth order, we 
can expand it as $f=f^{0}+\delta f$ where $f^{0}$ is the Planck function 
and $\delta f$ is the perturbation. It is convenient to describe the 
evolution of the perturbed distribution in terms of the brightness 
$\Delta({\mathbf{x}},\eta,\hat{n})$ defined as:
\begin{equation}
 \Delta=\frac{\int dp p^{3}\delta f}{\int dp p^{3}f^{0}}
\end{equation}
The space-time metric in comoving coordinates $x^{i}$ and conformal time  
$\eta$ for the  conformal-Newton gauge is given as (see e.g.
 Ma \& Bertschinger 1995 and discussion therein):
\begin{equation}
ds^{2}=a^{2}(\eta)[-(1+2\Psi)d\eta^{2}+(1-2\Phi)dx_{i}dx^{i}] 
\end{equation}
Here, $a(\eta)$ is the scale factor and $\Psi$, $\Phi$ are the two potentials 
characterising scalar perturbations in this gauge. 
The evolution of photons in this metric is then given by the Boltzmann 
equation for the brightness:
\begin{equation}
\dot{\Delta}+n_{i}\partial_{i}\Delta+n_{i}\partial_{i}\Psi+\dot{\Phi}=C[f]
\end{equation}
Here, over-dots denote derivative with respect to conformal time $\eta$. 
$C[f]$ is the collision term accounting for the scattering of photons 
with electrons. The linearised collision term for Thomson scattering 
(neglecting polarisation) is given as (see e.g. Hu \& White 1997):
\begin{equation}
C[f]=n_{e}\sigma_{T}\left(\Delta_{0}-\Delta+4{\mathbf{v}}_{e}\cdot\hat{n}+\frac{3}{2}n_{i}n_{j}\Pi_{ij}\right)
\end{equation}
Here, $\Delta_{0}\equiv\int\frac{d\Omega}{4\pi}\Delta$ denotes the 
isotropic part of $\Delta$ and $\Pi_{ij}$ is the photon anisotropic stress
 tensor which takes into account the angular dependence of Thomson
 scattering . It is given by: $\Pi_{ij}=\int\frac{d\Omega}{4\pi}\left
(n_{i}n_{j}-\frac{1}{3}\delta_{ij}\right)\Delta$.  
The nature of individual terms in $C[f]$ suggest that multipole moment 
expansion in terms of spherical harmonics can give a useful description. 
The details of such an expansion are given in the appendix. We finally 
arrive at a hierarchy of equations for moment $\Delta_{\ell m}$.
 We can see  that for a given $m$, each $l$-moment $\Delta_{\ell m}$ 
is coupled to an $l+1$ and an $l-1$ moment in the hierarchy. On the 
other hand, there is no coupling between different $m$ modes. This 
implies that if there is no source $S_{\ell m}$ (Appendix~A) 
for a given $m$ and if initial conditions are such that $\Delta_{\ell m}=0$ for that $m$, then, $\Delta_{\ell m}=0$ at all times 
even if other $m$ moments evolve. In the present analysis, our emphasis 
will be on studying the effect of scalar perturbations which correspond 
to $m=0$. For such perturbations  $\Pi_{ij}$ can be greatly simplified by 
using azimuthal symmetry about the axis of electron velocity (see e.g. 
Dodelson \& Jubas 1994). However our aim here is to study the generation 
of magnetic field from the evolution of coupled photon-baryon plasma. 
Since the generation of this field can explicitly break this symmetry, one 
should consider a more general expression for the anisotropic stress tensor. 

\subsection{Fluid equations for electrons and protons}
As discussed earlier, since the mean free paths of electrons and protons
are very small compared to astrophysical scales, we can describe their 
evolution accurately using continuity and Euler equations for an ideal fluid.  
In linear theory, the density field $\rho_{ e,p}({\mathbf{x}},\eta)$ 
is expanded as 
$\rho_{ e,p}({\mathbf{x}},\eta)=\bar\rho_{e,p}(\eta)(1+\delta_{e,p}({{\mathbf{x}},\eta}))$, 
where $\bar{\rho}$ is the unperturbed background density 
and $\delta$ is the fractional perturbation. In what follows 
quantities denoted with a bar on top are background unperturbed quantities. 
The continuity equations for each of the above species is given as (e.g. Ma
\& Bertschinger 1995):
\begin{equation}
\dot\delta_{e,p}+{\boldsymbol \nabla}\cdot{\bf v}_{e,p}-3\dot{\Phi}=0
\label{cont_eq}
\end{equation}

The corresponding  Euler equations are:
\begin{eqnarray}
{\dot{\mathbf{v}}_{e}}+\frac{\dot a}{a}{\mathbf{v}}_{e}=-\frac{{\boldsymbol \nabla} {P}_{e}}{\rho_{e}}-{\boldsymbol{\nabla}}\Psi-\frac{a e\mathbf{E}}{m_{ e}}-{ae \over m_e} {\bf v}_e \times {\bf B}+\left( \frac{\mathbf{v}_{\gamma}-\mathbf{v}_{e}}{\tau_{\gamma e}}\right)R+\frac{\mathbf{v}_{p}-\mathbf{v}_{e}}{\tau_{ep}} \label{euler_p}\\
\dot{\mathbf{v}}_{p}+\frac{\dot a}{a}{\mathbf{v}}_{p}=-\frac{{\boldsymbol \nabla} {P}_{p}}{\rho_{p}}-{\boldsymbol \nabla} \Psi+\frac{e\mathbf{E}}{m_{p}}+{ae \over m_p} {\bf v}_p\times {\bf B}+\frac{\mathbf{v}_{e}-\mathbf{v}_{p}}{\tau_{ep}}\left (\frac{m_{e}}{m_{p}}\right ) 
\label{euler_e}
\end{eqnarray}
Here, $\tau_{ep}$ is the (co-moving)  electron-proton collision time scale; 
$\tau_{\gamma e} = 1/(n_e \sigma_{\scriptscriptstyle T}a )$ is the 
photon-electron Thompson scattering time scale. ${\mathbf{E,B}}$ are the physical  electric and magnetic fields, $ R\equiv 4\rho_{\gamma}/3\rho_{e}$, and $P_{e,p}$ are the pressures which, for adiabatic fluids, can be written 
as $P\equiv P(\rho)$.  By taking curl of the Euler equations and using 
Maxwell's equations, we can get the evolution equation for the vorticities 
(${\mathbf{\Omega}}_{e,p}\equiv\nabla\times {\mathbf{v}}_{e,p}$) of the 
fluids as:
\begin{eqnarray}
\dot{{\mathbf{\Omega}}}_{e}+\frac{\dot{a}}{a}{\mathbf{\Omega}}_
{ e}&=&\frac{e}{m_{e}a}{d \over {d\eta}}(a^{2}\mathbf{B})
- {ae \over m_e} {\boldsymbol \nabla} \times ({\bf v}_e\times {\bf B})
+\left(\frac{{\mathbf{\Omega}}_{ \gamma}-{\mathbf{\Omega}}_{ e}}
{\tau_{\gamma e}}\right)R-\frac{\nabla^{2}{\mathbf{B}}}{4\pi n_e e\tau_{ep}} \nonumber \\
\dot{{\mathbf{\Omega}}}_{p}+\frac{\dot{a}}{a}{\mathbf{\Omega}}_{p}
&= & -\frac{e}{m_{p}a}{d \over {d\eta}}(a^{2}\mathbf{B})+ {ae \over m_p}{ \boldsymbol \nabla} \times ( {\bf v}_p\times {\bf B})
+ \frac{\nabla^{2}{\mathbf{B}}}{4\pi n_e e\tau_{ep}}\left(\frac{m_{e}}
{m_{\rss p}}\right) 
\label{ep_vort}
\end{eqnarray}

\section{Evolution equation for the magnetic field}
To arrive at the equation governing the evolution of magnetic field, we use 
the Euler equations for the charged fluids and Maxwell's equations 
(Appendix B). Subtracting Eq.~(\ref{euler_p}) from Eq.~(\ref{euler_e})
and using Maxwell's equations 
we first obtain the evolution of the current $\mathbf{J}$ :
\begin{equation}
{m_e \over e^2}{\partial \over \partial \eta}\left ({{\bf J} \over n_e}\right )+{\dot a \over a}{m_e \over e^2  n_e}{\bf J} =\frac{1}{n_{e}e}{\boldsymbol \nabla} 
P_{\rss e}+a{\mathbf{E}} + a ({\bf v}_e \times {\bf B})-\left(\frac{{\mathbf{v}}
_{\gamma}-{\mathbf{v}}_{e}}{\tau_{\gamma e}}\right)R {m_e \over e}-
\frac{m_e\mathbf{J}}{n_e e^2 \tau_{ep}}
\label{elec_field_evo}
\end{equation}
In the above equation we have neglected forces on the proton fluid due to 
pressure gradient and electric field since they are smaller than that for 
the electron fluid by the factor $m_{e}/m_{p}$. Taking curl of 
equation~(9) and using Maxwell's equations, we get the equation for 
the generation of magnetic fields:
\begin{equation}
{1 \over a} {\partial \over \partial \eta} (a^2 {\bf B})  = {m_e \over e^2} {\boldsymbol \nabla} \times {\partial \over \partial \eta} \left ({{\bf J} \over n_e} \right )+ {m_e \over e} {\boldsymbol \nabla} \times \left({{\boldsymbol \nabla} P_e \over \rho_e} \right ) -{\boldsymbol  \nabla} \times ({\bf v}_e \times {\bf B}) + {m_e \over e^2} {\boldsymbol \nabla} \times \left ({ {\bf J} \over n_e \tau_{ep} } \right ) + { m_e \over e} {\boldsymbol  \nabla} \times \left  ( { R ({\mathbf{v}}_{\gamma}-{\mathbf{v}}_e) \over \tau_{\gamma e} } \right ) - {\dot a \over a} {m_e \over e^2} {\boldsymbol  \nabla} \times \left ({ {\bf J} \over n_e } \right )
\label{mag_field_full}
\end{equation}
For studying the generation of magnetic fields in the early universe most of 
the terms in the above equation can be dropped. The first term on the right hand side can be shown to be negligible as compared to the term on the left hand side (see e.g. Widrow 2002). Similarly all the terms proportional to ${\bf B}$ 
can be neglected if one wishes to study the generation of magnetic fields
from zero magnetic field initial conditions. These terms can back-react once
the magnetic field is generated. We show later that the back-reaction terms
are negligible for the magnitude of the generated magnetic field. These 
considerations simplify the above equation to:
\begin{equation}
{1 \over a} {\partial \over \partial \eta} (a^2 {\bf B})  = {\bf S}({\bf x}, \eta)
\end{equation}
with 
\begin{equation}
{\bf S}({\bf x}, \eta) = {m_e \over e} {\boldsymbol \nabla} \times \left({{\boldsymbol \nabla} P_e \over \rho_e} \right ) + { m_e \over e} {\boldsymbol  \nabla} \times \left  ( { R ({\mathbf{v}}_{\gamma}-{\mathbf{v}}_e) \over \tau_{\gamma e} } \right )
\label{source_b}
\end{equation}
We now discuss the nature of these source terms of magnetic field generation in
first and second order in perturbation theory.

\subsection{Evaluation of the source term: Linear theory}
 The source term for any Fourier mode ${\mathbf{S}}({\mathbf{k}},\eta)$ can be 
simplified for the linear case. In this case, $R=4\bar\rho_{\gamma}/
(3\bar\rho_{e})$, and $\tau_{\gamma e}=1/(\bar n_e \sigma_
{\rss T})$, are unperturbed quantities and hence don't carry any spatial 
dependence. The first term of the right hand side of 
 Eq.~(\ref{source_b}) identically vanishes in this
case. The source term can then be written as:
\begin{equation}
{\mathbf{S}}({\mathbf{k}},\eta)=\frac{m_e R}{e\tau_{\gamma e}}({\mathbf{\Omega}}_{\gamma}-{\mathbf{\Omega}}_{e})
\end{equation}
Thus, we see that the source for the magnetic field in the plasma is the
 differential vorticity between electrons and photons. The vorticity 
equation for photons is essentially the Boltzmann moment equation 
for $l=1,m=1$ (Eq.~(\ref{boltz_eq})). We note that the source of photon
vorticity is $4v_{\ell1}/\tau_{\gamma e}\propto \Omega_{e}$. 
This implies that the only source which can excite any $l$-moment for 
$m=1$ is the electron fluid vorticity. The evolution equation for the 
electron fluid vorticity (Eq.~(\ref{ep_vort})) shows that the only sources 
of vorticity are the magnetic field and the photon vorticity. This means 
that if the vorticities were zero in the initial condition, as is the case 
with initial zero-vorticity conditions we consider here, none of these 
quantities can be generated for any scale in the linear regime. In particular 
we can conclude that no magnetic field is generated in linear order for 
scalar perturbations. It should be noted that using Eq.~(\ref{boltz_eq})~and~Eq.~(\ref{vor_sou}) 
allows us to follow modes at  which photons are free-streaming at any given 
epoch. Therefore the above conclusion holds for all scales larger than the 
scales at which electrons and protons can be treated as fluids. 

\subsection{Source term in the second order}
There are various terms which have to be included in going to second order
in perturbation theory. Second order terms can arise from treating metric 
perturbations to second order (Martinez-Gonzalez, Sanz \& Silk 1992) or 
by including the second order terms in the electron-photon scattering 
(Vishniac 1987, Jubas \& Dodelson 1995, Hu, Scott, \& Silk 1994). Vishniac 
(1987) adopted the simple procedure of including the spatial dependence of 
densities to include the second order effects. Detailed analyses (Jubas 
\& Dodelson 1995, Hu, Scott, \& Silk 1994) showed that Vishniac's procedure 
gives   the most important second order effect in the electron-proton 
scattering for sub-horizon scales. This allows us to study scales smaller
than the horizon at the last scattering surface, $H^{-1} \simeq 100 \, 
\rm Mpc$. At larger scales other second order effects from electron-photon 
scattering and the second order metric perturbations  might be 
comparable or dominate. 
We adopt Vishiniac's procedure here and obtain the second order term from
treating the spatial dependence of densities i.e. in $R$, $\tau_{\gamma e}$ 
and $\rho_e$ in the source term for magnetic field generation (Eq.~
(\ref{source_b})).  To get estimates of the  generated magnetic field 
we solve for the difference in photon and baryonic bulk velocity in the 
tight-coupling approximation. We argue below that the main contribution to 
the source $S({\bf k},\eta)$ for any scale comes from epochs at which the 
tight-coupling approximation is valid. The source term is evaluated in the 
tight-coupling approximation in Appendix~A and given by 
Eq.~(\ref{mag_gen_eq}). We argue there that for adiabatic evolution, the 
only term that can source the magnetic field generation is
given by Eq.~(\ref{mag_gen_eq1}). This allows us to solve the evolution of the 
generated magnetic field at any scale:
\begin{equation}
a^2 {\bf B}({\bf k},\eta) \simeq  {\bar R a  m_e \over 3 e} \int_0^{\eta}d\eta'{\dot a \over a} \int  d^3k' \delta_e({\bf k'}),\eta'){\bf k'}
 \times {\bf v}_e(({\bf k} - {\bf k'}),\eta')
\label{mag_field_gen2}
\end{equation}
Note that $a \bar R$ is independent of time. 
  Eq.~(\ref{mag_gen_eq1}) can 
be solved using linear theory evolution of density and velocity perturbations
for each scale from initial conditions at the time at which 
the scale is super-horizon to the epoch of recombination, $\eta_{\rm rec}$
(Eq.~(\ref{den_vel_evo})). At the 
epoch of recombination the photons decouple from the baryons and therefore
the source for magnetic field generation vanishes.  
We do not attempt an explicit solution here but seek an approximate 
understanding of the generated magnetic field.
 We first justify our use of the  tight-coupling 
approximation. As discussed in Appendix~A, for each scale the tight-coupling
approximation is valid for epochs before the Silk damping regime (Eq.~(\ref{delta_sol_silk})). 
From Eq.~(\ref{mag_field_gen2}), the magnetic field at a given scale ${\bf k}$ gets contribution from density and velocity perturbations at all scales. 
It should however be noted that if ${\bf k}$ corresponds to a scale
at which the density and velocity perturbations are in damping regime
 the source of magnetic field is also in the damping 
regime. (More precisely one is interested in the power spectrum of the 
magnetic field, which, from Eq.~(\ref{mag_field_gen2}),  is a four-point function containing density fields. 
For magnetic field at scale ${\bf k}$, the integrand of the source is 
$\propto P(k')P(|{\bf k} - {\bf k'}|)$, here $P(k)$ is the power spectrum 
of the density field;  $|{\bf k} - {\bf k'}| \simeq k$ if 
both $k$ and $k'$ are not in the damping or free-streaming regime.) This 
means that most of the contribution to the magnetic field comes from epochs 
at which the tight-coupling approximation is valid. More precisely much of 
the contribution to magnetic field at a scale $k$ comes from epochs 
$\eta \la \eta_d \simeq \omega_d^{-1}$ (Eq.~(\ref{damp_sc})). For scales that are not in the damping regime
at $\eta_{\rm rec}$ the upper limit of the integral in 
Eq.~(\ref{mag_field_gen2}) is $\eta_{\rm rec}$. For smaller scales, the upper
limit is $\simeq \eta_d$. Having identified some generic features of 
the magnetic field source terms we can give an order-of-magnitude estimate
of the generated field from Eq.~(\ref{mag_gen_eq1}). For all scales 
 a reasonable upper limit on the generated magnetic field for the
currently-favoured  $\rm \Lambda CDM$ model,  at  a scale $L \simeq k^{-1}$,  is:
\begin{equation}
(a B^2)(L, \eta_0) = (a B^2)(L, \eta_1) \simeq 10^{-30} \, {\rm G} \left (\delta_e(\eta_1) \over 10^{-3} \right ) \left (v_e(\eta_1, L) \over 10 \, {\rm km \, sec^{-1}} \right )
\left ( {10 \, {\rm Mpc} \over L } \right )
\label{gen_mag_field}
\end{equation}
Here $\eta_1 = \eta_{\rm rec}$ for scales that are not in Silk damping regime
at the epoch of last scattering and $\eta_1 \simeq \eta_d$ for 
smaller scales. In Eq.~(\ref{gen_mag_field}) we have used 
the fact that once the sources of generating magnetic fields vanish, 
the magnetic field evolves such that  $a^2 B$  remains constant (see e.g. 
Wasserman 1978). In Eq.~(\ref{gen_mag_field}), 
$v_e(L) \simeq (k P(k))^{1/2}$ (the 
matter power spectrum is given e.g. by Bardeen et~al. 1986)
 and $\delta_e(\eta_1)$ is the RMS of the density  field. One can 
compare the magnetic fields at different scales by evaluating the
 sources at the epoch of recombination $\eta_{\rm rec}$. As seen from Eq.~(\ref{den_vel_evo}) the density field is either non-evolving 
or in the oscillatory phase  for much of the period prior to recombination 
(except for modes  $\eta^{-1} \la  k \la \left( {\eta / \sqrt{3}} \right )^{-1}$ in the matter-dominated epoch). The velocity field however grows  
$\propto \eta$ for super-horizon scales in the radiation dominated era and 
for $k \la (\eta/\sqrt{3})^{-1}$ in the matter dominated era. Therefore if 
Eq.~(\ref{gen_mag_field}) is evaluated at $\eta_{\rm rec}$, 
 the small scale magnetic 
field is approximately smaller by a factor $(\eta_{\rm ent}/\eta_{\rm rec})$; 
here $\eta_{\rm ent} \simeq k^{-1}$ is the epoch of horizon entry of 
the mode $k$. Using this to write the sources of the magnetic field at 
$\eta_{\rm rec}$ in Eq.~(\ref{gen_mag_field}), it is seen that, for all 
scales,  $B(\eta_0) \la 10^{-30} \, \rm G$.

\section{Conclusion and discussion}
We have studied the possibility of generating magnetic fields 
during the evolution of the photon-baryon plasma in the pre-recombination
universe. For scalar perturbation in linear theory  magnetic field is 
not generated  at any scale; this includes  scales at which 
the photon-baryon coupling approximation breaks down.  We show that 
in the second order in perturbation theory 
 a small magnetic field is generated. The 
strength of the generated magnetic field is $\la  10^{-30}$ for
scales from $\simeq 100 \, \rm Mpc$ to sub-kpc at the present epoch. 

In Eq.~(\ref{mag_field_full}), we have neglected several terms which could 
back-react on the generated magnetic field. It can be readily seen that,
for the strength of the generated field, these terms are always much
smaller than the source term of the magnetic field. 
And therefore we were  justified
in neglecting those terms for studying  the generation of  magnetic field.  As discussed above magnetic fields at small scales are frozen 
in the plasma from  epochs $\simeq \omega_d^{-1}$. It could be asked whether
the radiative viscosity prior to the recombination can damp these fields. 
The maximum length scale damped by pre-recombination radiative viscosity is 
$\propto B$ (Jedamzik, Katalini{\' c}, \& Olinto 1998, Subramanian \& Barrow 1998). For the small magnetic fields we obtain, the maximum 
scale of dissipation   can be 
shown to be  much smaller than  any relevant length scales.

\appendix

\section{Photon Boltzman Equation and Initial Conditions}
In this appendix we discuss the complete spherical harmonic moment expansion of the Boltzmann equation without assuming azimuthal symmetry of $\Delta$. The assumption of azimuthal symmetry precludes the existence of vortical modes for the
velocity fields of photons and electrons. Hence if we are looking for the generation of magnetic fields in the linear order, we have to relax this assumption since a possible presence of magnetic fields violates azimuthal symmetry. The  notations used in this section are self contained. \\
The Fourier transformed Boltzmann equation can be written as:
\begin{equation}
\dot{\Delta}+ik\mu\Delta+ik\mu\Psi+\dot{\Phi}=\frac{1}{\tau_{\gamma e}}\left(\Delta_{0}-\Delta+4{\mathbf{v}}_{e}\cdot\hat{n}+\frac{3}{2}n_{i}n_{j}\Pi_{ij}\right)
\label{delta_eq}
\end{equation}
The photon brightness function $\Delta$ can be expanded in terms of the scalar spherical harmonics $Y_{\ell m}$ as:
\begin{equation} 
\Delta(\hat{n})=\sum \sqrt{\frac{4\pi}{2\ell+1}}\Delta_{\ell m}Y_{\ell m}(\hat{n})
\end{equation}
where, the coefficients $\Delta_{\ell m}$ are given by the inverse relation,
\begin{equation}
\Delta_{\ell m}=\sqrt{\frac{2\ell+1}{4\pi}}\int{d\Omega} Y^{*}_{\ell m}(\hat{n})\Delta(\hat{n})
\end{equation}
The photon fluid variables like over-density $\delta_{\gamma}$ and velocity ${\mathbf{v}}_{\gamma}$ are then given by:
\begin{eqnarray}
\delta_{\gamma}\equiv\int\frac{d\Omega}{4\pi}\Delta=\Delta_{00}\\
\hat{k}\cdot {\mathbf{v}}_{\gamma}=\frac{\Delta_{10}}{4}\\
\Omega_{\gamma}=|\hat{k}\times {\mathbf{v}}_{\gamma}|=\frac{\Delta_{11}}{4}\\
\end{eqnarray}
By substituting the expansion for $\Delta$ in Eq.~(\ref{delta_eq}) and using 
the familiar properties of spherical harmonics we arrive at the following 
hierarchy of equations for the evolution of the moments $\Delta_{\ell m}$. 
For details of such an expansion we refer to the paper (Hu and White 1997). 
\begin{equation}
{\dot{\Delta}}_{\ell m}+ik\frac{A_{\ell,m}}{(2\ell-1)}\Delta_{\ell-1,m}+ik\frac{A_{\ell+1,m}}{(2\ell+3)}\Delta_{\ell+1,m}+\frac{\Delta_{ \ell m}}{\tau_{\gamma e}}=S_{\ell m}
\label{boltz_eq}
\end{equation}
Here, $A_{\ell m}=\sqrt{\ell^{2}-m^{2}}$. The source $S_{\ell m}$ is given as:
\begin{equation}
S_{\ell m}=\left(\frac{\Delta_{\ell m}}{\tau_{\gamma e}}+4\dot{\Phi}\right)\delta_{\ell 0}\delta_{m0}+\left(\frac{4v_{\ell m}}{\tau_{\gamma e}}-k\Psi\delta_{m0}\right)\delta_{\ell 1}+\frac{1}{10}\Delta_{\ell m}\delta_{\ell 2}
\label{vor_sou}
\end{equation}
In the above equations, the coefficients $v_{lm}$ are the coefficients in 
the multipole expansion of the ${\mathbf{v}}_{e}\cdot\hat{n}$ term such 
that ${\mathbf{v}}_{e}\cdot\hat{n}=\sum v_{\ell m}Y_{\ell m}\delta_{\ell 1}$.
The vorticity of the photon fluid is tracked by the evolution of the 
$\ell =1,m=1$ moment $\Delta_{11}$. We notice from Eq~(\ref{vor_sou}), that 
the source $S_{11} = 4v_{11}/\tau_{\gamma e}$  This is the only source for 
the evolution of the $m=1$ moment. The first two moment equations in the 
hierarchy give the familiar continuity and Euler equations for photons:
\begin {equation}
{\dot{\delta}}_{\gamma}+\frac{4i}{3}{\mathbf{k}}\cdot {\mathbf{v}}_
{\gamma}-4\dot{\Phi}=0  
\label{phot_cont}     
\end{equation}
\begin{equation}
\dot{{\mathbf{v}}}_{\gamma}+i{\mathbf{k}}\frac{\delta_{\gamma}}{4}+
{\mathbf{\Pi}}-i{\mathbf{k}}\Psi=\frac{{\mathbf{v}}_{e}-{\mathbf{v}}_
{\gamma}}{\tau_{\gamma e}}
\label{pho_vel}
\end{equation}
In the above equation, ${\mathbf{\Pi}}_{i}=\frac{3}{4}ik_{j}\Pi_{ij}$.

\subsection{Initial Conditions: Tight-coupling approximation}
The electron-proton plasma recombines at a redshift $z_{\rm rec} \simeq 10^3$
(see e.g. Peebles 1993). 
At any epoch in the universe before recombination, $\eta \la \eta_{\rm rec} 
\simeq 2 H_0^{-1}(1+z_{\rm rec})^{-1/2}/\Omega_m^{1/2}$, there are roughly 
five physically relevant length scales:
(a) Super-horizon scale, $k \la \eta^{-1}$ (b) scales that are sub-horizon 
but larger than the sound Horizon, $\eta/\sqrt{3} \ga k \ga \eta^{-1}$. At 
these scales the evolution of velocity fields is determined by gravitational 
potentials. (c) scales  smaller than the sound horizon scale but larger than
the Silk damping  scale, $\eta/\sqrt{3} \la k \la k_{\rm silk}$. At these 
scales the baryon velocity evolution is determined by both 
 gravitational potentials and  the photon pressure, (d) 
scales that  are in the damping regime but larger than the scales at which 
photon free-stream, $k_{\rm silk} \la k \la k_{\rm fs}$, $k_{\rm fs} \simeq (2 \, \rm Mpc)^{-1} (10^3/(1+z))^{-2}$. The densities and 
velocities of baryons decay exponentially in this
regime (see e.g. Peebles 1980) and (e) $k \ga k_{\rm fs}$, at 
these scales photons are free-streaming and therefore 
photons and baryons cannot 
be treated as  coupled fluids. During   the evolution in the 
expanding universe before recombination,  the electron
velocity and density perturbations at most   scales  first 
pass through the Silk damping regime before reaching this phase. 
Therefore during this phase $\delta_e, v_e \simeq 0$. The only
exception to this occurs around the epoch of recombination when
the free-streaming length increases very rapidly. As the sources
of magnetic field generation are nearly zero in this regime, the 
dynamics of plasma at these scales play an unimportant role
for our study. In the evolution in linear theory all scales 
 undergo either some or all of these phases of evolution. 

Initial condition for each mode is set outside the horizon. 
Up to  phase (c) discussed  above, $k \ll k_{\rm fs}$. During 
this  phase the photons are tightly coupled to the baryons and 
this greatly simplifies the problem (Peebles \& Yu 1970, Peebles 1980, Hu \& 
Sugiyama 1995). In this approximation, to zeroth order in $\tau_{\gamma e}$:
$\mathbf{v}_{\gamma} = {\mathbf{v}}_{e}$; and  $\Pi^{ij} = 0$.
 Also to this order in $\tau_{\gamma e}$: $\delta_e =   3/4 \delta_\gamma$;
this can be readily obtained by subtracting   the electron 
continuity equation from the photon continuity equation 
(Eqs.~(\ref{phot_cont})~and~(\ref{cont_eq})).  To solve for the difference 
between electron and photon bulk velocity we need to expand to the 
first order in $\tau_{\gamma e}$. To this order, from Eq.~(\ref{pho_vel}) (Peebles \& Yu 1970, Peebles 1980):
\begin{equation}
{\bf v}_{\gamma} -{\bf v}_{e}= \tau_{\gamma e}\left ({\partial {\bf v}_{\gamma}^i \over \partial \eta } + {1 \over 4}{{\boldsymbol \nabla }\delta_\gamma } + 
 {\boldsymbol\nabla}{\bf \Pi}  + {\boldsymbol \nabla}\Psi \right )
\label{tcoupl}
\end{equation}
Here the quantities in the bracket on the right hand side are to be 
evaluated to the zeroth order in $\tau_{\gamma e}$. 
 Eq.~(\ref{tcoupl}) 
along with the evolution of electron velocity (Eq.~(\ref{euler_e})) 
and the difference of electron and proton velocities  (Eq.~(\ref{elec_field_evo})) can be used to give
the following expression for the electric field in the tight coupling approximation:
\begin{equation}
a {\bf E}({\bf x},t) = {m_e \over e} \left ({\dot a \over a} R {\bf v}_e - {{\boldsymbol  \nabla} p_e \over \rho_e} - {1 \over 4}R{ \boldsymbol\nabla} \delta_\gamma \right )
\label{elec_fi}
\end{equation}
In deriving Eq.~(\ref{elec_fi}) all terms proportional to the magnetic field
were dropped as they cannot act as  sources for generating magnetic field.
Taking the curl of this equation and using  Maxwell's equation (Eq.~(\ref{mag_gen})) one obtains the equation for magnetic field generation:
\begin{equation}
{1 \over a} {\partial \over \partial \eta} (a^2 {\bf B}) =  {m_e \over e} { \boldsymbol \nabla} \times  \left (-{\dot a \over a} R {\bf v}_e + {{\boldsymbol  \nabla} p_e \over \rho_e} + {1 \over 4}R{\boldsymbol  \nabla} \delta_\gamma \right )
\label{mag_gen_eq}
\end{equation}
This equation verifies the discussion above that the source of magnetic field
generation is electron vorticity in the linear theory. With
non-vortical initial conditions,  Eq.~(\ref{mag_gen_eq})
shows that all the sources of magnetic field generation are zero in the 
linear perturbation theory. We wish to consider the second 
order effect by considering the spatial dependence of densities. This
gives: $R = \bar R (\delta_e - \delta_\gamma) \simeq -1/3 \bar R \delta_e$ 
in the tight coupling approximation, as   $\delta_e =   3/4 \delta_\gamma$ during adiabatic expansion  (see e.g. Peebles 1980). 
In second order, the second term
on the right hand side is the Biermann battery term. In the adiabatic initial
condition we consider here, $p = \rho^\gamma$ with $\gamma = 5/3$.  And 
the source term $\propto \nabla \delta_e \times \nabla \delta_e = 0$
and  therefore  in this limit the Biermann  battery term doesn't contribute.
It should be noted that the plasma evolves adiabatically only for scales that
are not affected  by Silk damping (see below). However as the densities
and velocities  
damp in this regime one doesn't expect much contribution from these scales.
Biermann battery term can also contribute for initial conditions different
from the adiabatic initial conditions. The third term on the 
right hand side also   vanishes even  in the second order 
in the tightly-coupled regime. Therefore the only source of magnetic field
generation is the first term on the right hand side of Eq.~(\ref{mag_gen_eq}).
Eq.~(\ref{mag_gen_eq}) can therefore be simplified to:
\begin{equation}
 {1 \over a} {\partial \over \partial \eta} (a^2 {\bf B}) =  {\bar R m_e \over 3 e} {\boldsymbol \nabla} \times  \left ({\dot a \over a} \delta_e  {\bf v}_e \right )
\label{mag_gen_eq1}
\end{equation}  
This equation can be used to get an order-of-magnitude estimate of the 
generated magnetic field.

 Eq.~(\ref{tcoupl})  can be used  to calculate, to the zeroth order, 
 the evolution equation of electron velocity field (Peebles \& Yu 1970).
 This  equation along with the continuity equation (Eq.~(\ref{cont_eq})) and 
${\bf \nabla .E} = 0$, Eq.~(\ref{diver_e}),
 and dropping all terms proportional to ${\bf B}$, gives:
\begin{equation}
\ddot \delta_e = -{\dot a \over a}{\dot \delta_e \over (1+R)} - {k^2 p_e  \over \rho_e (1+R)} - k^2 \Psi -{R \over 4 (1+R)} k^2 \delta_\gamma+ {3\dot a \over a} {\dot \Phi \over (1+R)} + 3  \ddot \Phi
\label{elec_evol}
\end{equation}
This equation can be solved along with the evolution equation of $\delta_\gamma$ by WKB approximation and these solutions can be matched to large 
scale solutions (Hu \& Sugiyama 1995). We discuss 
here approximate solutions at  different epochs. First we discuss solutions 
during phase (c) of the evolution. 
 We note that all the terms on the right hand 
side except for the $\delta_\gamma$ term are smaller as compared to 
this term for scales smaller than the 
 sound horizon scale $\simeq 1/\sqrt{3} \eta$. The electron pressure
 is always negligible as compared to the photon pressure  in the 
pre-recombination universe. 
With these simplification and bearing in mind that $1/R \ll 1$ during the 
evolution, Eq.~(\ref{elec_evol}) is solved to give:
\begin{equation}
\delta_e({\bf k}, \eta)= A({\bf k}) \cos\left (\int_0^\eta \omega_o d\eta'\right )
\label{delta_sol}
\end{equation}
Here we have only retained the solution compatible with adiabatic initial
conditions (see e.g. Hu \& Sugiyama 1995) and 
\begin{equation}
\omega_o = {k \over \sqrt{3 (1+ 1/R)}}
\end{equation}
In phase (d) of the evolution of the plasma, the tight-coupling approximation
breaks down and the photon-slip which damps perturbations (Silk damping)  
must be taken into account. The  solution 
including the Silk damping is (see e.g. Peebles 1980):
 \begin{equation}
\delta_e({\bf k}, \eta) = A({\bf k}) \cos(\omega_o \eta) \exp(-\omega_d \eta)
\label{delta_sol_silk}
\end{equation}
Here, 
\begin{equation}
\omega_d \simeq  {2 k^2 \tau_{\gamma e} \over 15 } 
\label{damp_sc}
\end{equation}
The silk damping scale, at any epoch, can be obtained from this 
expression:  $k_{\rm silk} \simeq (15/(2\tau_{\gamma e}\eta))^{1/2} 
\simeq (4 \, {\rm Mpc})^{-1} ((1+z)/10^3)^{-5/4}$ in the matter-dominated 
era. 
The velocity field in the linear evolution remains non-vortical, and hence
can be found from the continuity equation (Eq.~(\ref{cont_eq})). It should 
be noted that  solutions for the baryon density and velocity fields 
differ from the corresponding quantities for the electrons only by replacing 
$R$ defined here as $R' = 4\rho_\gamma/(3\rho_b)$ (see e.g. Hu \& 
Sugiyama 1995).  As $R' \gg  1$ for the evolution of the plasma in the 
pre-recombination universe, for baryonic densities compatible with 
primordial nucleosynthesis, the baryon and electron quantities can be used 
interchangeably in evaluating the second order expression above.

The evolution of electron density and velocity in the oscillatory regime
and   the super-horizon solutions,  prior to the 
epoch of recombination, can be summarized as (for solutions
at super-horizon scales in this conformal-Newton
gauge see e.g. Ma \& Bertschinger 1995):
\begin{eqnarray}
\delta_e & \propto  & \Psi = \hbox{constant} \quad \hbox{for} \quad k \la \eta^{-1} \quad ( \hbox{RD and MD}) \nonumber \\
\delta_e & = & \hbox{oscillatory} \quad \hbox{for} \quad  k \ga \left({\eta \over \sqrt{3}}\right )^{-1}\quad (\hbox{RD and MD})  \quad \hbox{and for} \quad  k \ga \eta^{-1} \quad (\hbox{RD}) \nonumber \\
\delta_e & \propto &  \eta^2 \quad \hbox{for} \quad  \eta^{-1} \la  k \la \left( {\eta \over \sqrt{3}} \right )^{-1} \quad (\hbox{MD}) \nonumber  \\
v_e & \propto  & k \Psi \eta \quad \hbox{for} \quad k \la \eta^{-1} \quad ( \hbox{RD and MD}) \quad \hbox{and for} \quad \left ({\eta\over \sqrt{3}} \right )^{-1} \ga  k \ga \eta^{-1} \quad (\hbox{MD}) \nonumber  \\
v_e &  =  &  \hbox{oscillatory}  \quad \hbox{for} \quad k \ga \left ({\eta \over \sqrt{3}} \right )^{-1} \quad ( \hbox{RD and MD}) 
\label{den_vel_evo}
\end{eqnarray}
Here ${\rm RD}$ and ${\rm MD}$ correspond to radiation and matter dominated 
epochs,  respectively.

\section{Maxwell's equations for FRW background}
The Maxwell's equations for the FRW metric  in terms of physical fields, $\mathbf{E,B,J}$ are as follows:
\begin{eqnarray}
{\boldsymbol \nabla}\times(a^{2}{\mathbf{B}})=4\pi a^{3}{\mathbf{J}}+\frac{\partial(a^{2}{\mathbf{E}})}{\partial \eta} \label{curl_b}\\
{\boldsymbol\nabla}\cdot{\mathbf{B}}=0 \\
{\boldsymbol \nabla}\times(a^{2}{\mathbf{E}})=-\frac{\partial(a^{2}{\mathbf{B}})}{\partial \tau} \label{mag_gen}\\
{\boldsymbol \nabla} \cdot{(a^{2}{\mathbf{E}})}=4\pi a^{3}e(n_{ p}-n_{e}) 
\label{e_div}
\end{eqnarray}
The current $\mathbf{J}$ is written in terms of fluid quantities as:
\begin{equation}
{\mathbf{J}}=e(n_{ p}{\mathbf{v}}_{ p}-n_{ e}{\mathbf{v}}_{e})
\label{curre}
\end{equation}
Here, $n_{e,p}$ are the electronic and protonic number densities which are assumed to be equal to the lowest order i.e ${\bar{n}}_{e}={\bar{n}}_{ p}=n$. From Eq.~(\ref{curl_b}) it follows that ${\bf \nabla.J} = 0$ 
if the second term can be neglected, which is the case here (see e.g. Parker 1979).  Eq.~(\ref{cont_eq}) along with Eq.~(\ref{curre}) then shows that:
\begin{equation}
{\boldsymbol \nabla.{\bf E}} = 0
\label{diver_e}
\end{equation}
in the linear theory. 

\section{Acknowledgement}
One of us (SKS) would like to thank K. Subramanian and A.  Reisenegger  for 
many useful discussions.


\begin{thebibliography}{}
\bibitem[]{} {Bardeen}, J.~M., {Bond}, J.~R., {Kaiser}, N. \&  {Szalay}, A.~S. 1986, ApJ, 304, 15
\bibitem[]{}  Beck R., Brandenburg A., Moss D., Shukurov A. M., Sokoloff
D. D., 1996, Ann. Rev. Astron. Astrophys., 34, 155
\bibitem[]{} Berezhiani, Z.   \& Dolgov, A. D. 2003, astro-ph/0305595
\bibitem[]{} Brandenburg, A. \& Subramanian, K., 2004, astro-ph/0405052 
\bibitem[]{}{Dodelson}, S. \&  {Jubas}, J.~M. 1995, ApJ, 439, 503
\bibitem[Freedman et al.(2001)]{2001ApJ...553...47F} Freedman, W.~L.~et 
al.\ 2001, \apj, 553, 47
\bibitem[]{MG03}  Giovannini, M, 2004, Int.J.Mod.Phys., D13, 391 
\bibitem[]{rubgras}  Grasso D., Rubinstein H. R., 2001, Phys. Rep., 348, 161
\bibitem[Harrison(1970)]{1970MNRAS.147..279H} Harrison, E.~R.\ 1970, 
\mnras, 147, 279 
\bibitem[]{} Hogan, C. 2000, astro-ph/0005380
\bibitem[Hu \& Sugiyama(1995)]{1995ApJ...444..489H} Hu, W.~\& Sugiyama, N.\ 
1995, \apj, 444, 489 
\bibitem[]{} Hu, W. \&  {White}, M. 1997, Phys. Rev D, 56, 596
\bibitem[]{} Hu, W.,  Scott, D. \& Silk, J. 1994, Phys. Rev. D, 49, 648
\bibitem[Jedamzik, Katalini{\' c}, \& Olinto(1998)]{1998PhRvD..57.3264J} 
Jedamzik, K., Katalini{\' c}, V., \& Olinto, A.~V.\ 1998, \prd, 57, 3264 
\bibitem[Kulsrud, Cen, Ostriker, \& Ryu(1997)]{1997ApJ...480..481K} 
Kulsrud, R.~M., Cen, R., Ostriker, J.~P., \& Ryu, D.\ 1997, \apj, 480, 481 
\bibitem[]{} Ma, C. and Bertschinger, E. 1995, ApJ, 455, 7
\bibitem[Parker(1979)]{1979cmft.book.....P} Parker, E.~N.\ 1979, Oxford, 
Clarendon Press; New York, Oxford University Press, 1979, 858 p.
\bibitem[]{} Peebles, P.~J.~E.\ 1993, 
Principles of Physical Cosmology, Princeton University Press
\bibitem[]{} Peebles, P.~J.~E.\ 1980, The Large Scale Structure of the Universe, Princeton University Press
\bibitem[]{} Peebles, P.~J.~E. and {Yu}, J.~T. 1970, ApJ, 162, 815
\bibitem[]{} Peiris, H.~V. et~al. 2003, ApJS, 148, 213
\bibitem[Perlmutter et al.(1999)]{1999ApJ...517..565P} Perlmutter, S.~et 
al.\ 1999, \apj, 517, 565
\bibitem[Ratra(1992)]{1992ApJ...391L...1R} Ratra, B.\ 1992, \apjl, 391, L1
\bibitem[]{} Riess, A.~G.~et al.\ 2004  astro-ph/0402512
\bibitem[Riess et al.(1998)]{1998AJ....116.1009R} Riess, A.~G.~et al.\ 
1998, \aj, 116, 1009
\bibitem[]{}  Ruzmaikin A. A., Shukurov A. M., Sokoloff D. D., 1988, {\it %
Magnetic Fields of Galaxies}, Kluwer, Dordrecht (1988)
\bibitem[]{} Spergel, D. N. et al. 2003, ApJS, 148, 175
\bibitem[Subramanian \& Barrow(1998)]{1998PhRvD..58h3502S} Subramanian, 
K.~\& Barrow, J.~D.\ 1998, \prd, 58, 83502 
\bibitem[Subramanian, Narasimha, \& Chitre(1994)]{1994MNRAS.271L..15S} 
Subramanian, K., Narasimha, D., \& Chitre, S.~M.\ 1994, \mnras, 271, L15 
\bibitem[]{shukurov04} Shukurov, A., 2004, Introduction to galactic dynamos,
In Mathematical aspects of natural dynamos, Ed. E. Dormy,
Kluwer Acad. Publ., Dordrecht.
\bibitem[]{} Tonry, J. L. et al.~2003, \apj, 594, 1
\bibitem[Turner \& Widrow(1988)]{1988PhRvD..37.2743T} Turner, M.~S.~\& Widrow, L.~M.\ 1988, \prd, 37, 2743
\bibitem[Tytler, O'Meara, Suzuki, \& Lubin(2000)]{2000PhR...333..409T} 
Tytler, D., O'Meara, J.~M., Suzuki, N., \& Lubin, D.\ 2000, \physrep, 333, 
409 
\bibitem[]{} Vishniac, E.~T. 1987, ApJ, 322, 597 
\bibitem[Wasserman(1978)]{1978ApJ...224..337W} Wasserman, I.\ 1978, \apj, 
224, 337
\bibitem[Zeldovich, Ruzmaikin, \& Sokolov(1983)]{1983flma....3.....Z} 
Zeldovich, I.~B., Ruzmaikin, A.~A., \& Sokolov, D.~D.\ 1983, New York, 
Gordon and Breach Science Publishers (The Fluid Mechanics of Astrophysics 
and Geophysics.~Volume 3), 1983, 381 p.
\end{thebibliography}
\end{document}